\documentclass[prc,preprintnumbers,showpacs,showkeys,superscriptaddress,10pt]{revtex4} 

\usepackage{enumerate} 

\usepackage{amsmath}  

\usepackage{amsthm}  

\usepackage{amsfonts}  

\usepackage{amssymb} 

\usepackage{amsxtra}

\usepackage{amssymb}

\usepackage{delarray} 

\usepackage{graphicx}

\usepackage{dcolumn}  

\usepackage{epsfig}  

\usepackage{float}

\usepackage{hhline}

\usepackage{color}

\restylefloat{figure}

\begin{document}

\title{Separable form of low-momentum realistic NN interaction}

\author{P. Grygorov}

\affiliation{Institut f\"{u}r Theoretische Physik,Universit\"{a}t T\"{u}bingen, 
D-72076 T\"{u}bingen, Germany}
\author{E.N.E. van Dalen}

\affiliation{Institut f\"{u}r Theoretische Physik,Universit\"{a}t T\"{u}bingen, 
D-72076 T\"{u}bingen, Germany}
\author{J. Margueron}
\affiliation{Institut de Physique Nucl\'eaire,
Universit\'e Paris-Sud, F-91406 Orsay CEDEX, France}

\author{H. M\"{u}ther}

\affiliation{Institut f\"{u}r Theoretische Physik,Universit\"{a}t T\"{u}bingen, 
D-72076 T\"{u}bingen, Germany}

\begin{abstract}
The low-momentum interaction $V_{\text{low-k}}$ derived from realistic models of
the nucleon-nucleon interaction is presented in a separable form. 
This separable force is supported by a 
contact interaction in order to achieve the saturation properties of
symmetric nuclear matter. 
Bulk properties of nuclear matter and finite nuclei are investigated 
for the separable form of $V_{\text{low-k}}$ and two different parameterizations
of the contact term.  
The accuracy of the separable force in Hartree-Fock calculations 
with respect to the original interaction $V_{\text{low-k}}$ is discussed.  
For a cutoff parameter $\Lambda$ of 2 fm$^{-1}$ a representation by a rank 2
separable force yields a sufficient accuracy, while higher ranks are required
for larger cut-off parameters. 
The resulting separable force is parameterized in a simple way to allow for an
easy application in other nuclear structure calculations.

\end{abstract}




\maketitle

\section{Introduction}
\label{sec:Introduction}

The evaluation of bulk properties of finite nuclei and nuclear matter starting from
realistic models of nucleon-nucleon (NN) interaction is a major challenge
in modern nuclear physics. Since the exact form of the interaction resulted from 
the underlying theory of the strong interaction remains 
unknown, one usually has deal with realistic models developed so as to fit experimental 
data for free nucleon-nucleon scattering up to the threshold for pion production 
and properties of the deuteron \cite{Machleidt96}-\cite{Entem03}. 
It was done by obtaining a best fit for a large number of adjustable parameters 
using several thousands experimental points so that there exists several quite different
potential models commonly used.
A general feature of such realistic interactions is strong short-range and tensor components,
which cannot be handled within the standard perturbation theory. 
There have been suggested different approaches in order to overcome this problem:
Bethe-Brueckner-Goldstone expansion \cite{Baldo99}, correlated basis functions \cite{Feenberg69}, 
quantum Monte Carlo \cite{Wiringa00},
self-consistent Green's function theory (see, e.g., Ref. \cite{Dickhoff05}).
These methods were successfully applied to describe bulk properties of nuclear matter \cite{Goegelein09}, 
pairing gap of nucleons \cite{Fabrocini08}, weak response \cite{FarinaPhD} and shear viscosity
of nuclear matter \cite{Benhar09}.
However these approaches remain very complex to be applied 
 {directly}
to a description of finite nuclei, as well as inhomogeneous nuclear
matter, also known as pasta phase, which exists in the inner crust of neutron stars.
 {Alternatively, they have been combined either to phenomenological approaches 
through a local density approximation~\cite{Baldo2004}, or as an input for a density 
functional approach~\cite{Baldo2008}. In these approaches, adjustable parameters need
however to be determined.}

Besides the realistic interactions, various phenomenological models have been 
developed, 
 {such as the Skyrme interaction~\cite{Skyrme59}, and}
adjusted to describe the experimental 
data for the ground states of finite nuclei 
and the empirical saturation point of symmetric nuclear matter. 
A simple parameterization of such phenomenological forces 
through the local single-particle densities allows
a simple solution of the Hartree-Fock (HF) equations~\cite{Vautherin1972}. 
Finally, these models have been successfully 
used for predictions of equations of state (EoS) 
of nuclear matter and description of pasta phase within the Wigner-Seitz 
(WS) cell approximation~\cite{Negele73}.
 {In neutron stars, these models are extrapolated far from the condition where it has 
been adjusted and might in some cases become unstable~\cite{Margueron2002}.
The instabilities of these models could however be corrected such as it reproduces the features of a
G-matrix in nuclear matter~\cite{Margueron2009}.}
%

An alternative method, which is based on realistic NN interactions
and allows us to perform Hartree-Fock calculations similarly 
 {to}
the phenomenological forces is the low-momentum interaction $V_{\text{low-k}}$. 
The basic idea of $V_{\text{low-k}}$ is to separate the predictions 
for correlations at low momenta, which are constrained by the NN scattering matrix below the pion threshold,
from the high-momentum components, which may strongly depend on the underlying model
of realistic NN interaction. By introducing a cutoff $\Lambda$ in momentum
space, one separates the Hilbert space into a low-momentum and a high-momentum
part. The renormalization technique (see, e.g.,
\cite{Bogner01,Lee80,Okubo54,Bogner08,Suzuki82}) 
determines an effective Hamiltonian, which
must be diagonalized within the model space (below the cutoff).  
With the cutoff in the range of $\Lambda$ = 2 fm$^{-1}$ $V_{\text{low-k}}$ becomes
model independent, and reproduces the deuteron binding energy,
low-energy phase shifts, and half-on-shell $T$ matrix with the same accuracy
as the initial realistic interaction. This model independence demonstrates
that the low-momentum physics does not depend on details of the high-momentum
dynamics.

In spite of its obvious advantages $V_{\text{low-k}}$ potential still remains
 {a}
quite complicated object. On the one hand, it is nonlocal and therefore is represented 
as a matrix element in momentum space
for each partial wave channel. This nonlocality increases the computational time
 {in Hartree-Fock iteractions, and prevents the use of 
$V_{\text{low-k}}$ if the number of nucleons is too large, such as in the
Wigner-Seitz cells present in the crust of neutron stars for instance~\cite{Negele73}.}
%
On the other hand, the renormalization technique used to produce $V_{\text{low-k}}$
seems not to be trivial.
 {The resulting interaction is given as a matrix table which
is not an easy-to-use form and prevents this potential to be popular.}
%
A possible way out is to find a separable representation of $V_{\text{low-k}}$,
since it significantly simplifies many-body calculation \cite{Baldo88,Tian09}.
Moreover recent calculations of triton binding energies demonstrate the $V_{\text{low-k}}$
can be very good approximated by a low-rank separable force for low values of the cutoff
$\Lambda$ \cite{Kamada06}.
We investigate the separability of $V_{\text{low-k}}$ by using the diagonalization
of the matrix in momentum space for each partial wave channel. It allows us
to find a low rank separable form of $V_{\text{low-k}}$, which can be used in 
HF calculations of nuclear matter as well as finite nuclei.

The $V_{\text{low-k}}$ Hartree-Fock calculations 
demonstrate a monotonic increase of 
 {the}
binding energy of symmetric nuclear matter
as a function of the nucleon density, thus it cannot reproduce the empirical saturation
point \cite{Bozek06,Kuckei03}. Therefore we supplement $V_{\text{low-k}}$ 
by a simple density-dependent 
contact term, which accounts for a three-body correlations. This contact term is
adjusted to reproduce the saturation property of symmetric matter.

The paper is organized as follows. In the next Section \ref{sec:NN-interaction} 
we discuss the model 
space technique used to produce $V_{\text{low-k}}$ and outline the procedure to
determine the separable representation. In the last Section \ref{sec:Results} we sum up all
results and suggest a simple fit for separable representation of $V_{\text{low-k}}$
as well as two different parameterizations of the contact term, adjusted for
the fitted potential.

\section{Model of the NN Interaction}
\label{sec:NN-interaction}
The main idea of $V_{\text{low-k}}$ interaction is to disentangle 
the low-momentum or long-range part of a realistic NN interaction, 
which is fairly well described in terms of meson-exchange,
from the high-momentum or short-range part 
where quark degrees of freedom are getting important.
In other words, one defines a model space, which accounts for the 
low-momentum degrees of freedom and renormalizes the effective Hamiltonian
for this low-momentum regime in order to account for the effects
of the high-momentum components, 
which are integrated out.

In practice the $V_{\text{low-k}}$ interaction can be derived either
using model space methods (such as Lee-Suzuki \cite{Lee80} or Okubo \cite{Okubo54}) 
or through an renormalization group treatment \cite{Bogner01}. Both approaches
are essentially equivalent
and lead to the same energy-independent potential \cite{Bogner08}. 
In the following we will use the model space technique to disentangle
these parts based on the unitary model operator approach (UMOA). This approach
has frequently been described in the literature
\cite{Suzuki82,Bozek06,Fujii04}.
Therefore we will restrict the presentation only to basic equations,
which will define the nomenclature. 
  
To determine the model space, the low-momentum subspace of Hilbert space, one defines a projection 
operator $\hat{P}$, which projects
onto this model space. 
The complement of the subspace will be defined by the projection operator 
$\hat Q$, in such a way
 {that}
the whole space 
%
 {is}
covered by these two 
operators. Thus they satisfy the following relations $\hat P + \hat Q =1$, $\hat P^2=\hat P$, $\hat Q^2=\hat Q$, 
$\hat P \hat Q=0=\hat Q \hat P$.
The unitary model operator approach defines a unitary transformation $\hat U$ in such a way
that the transformed Hamiltonian
does not couple the $\hat P$ and $\hat Q$ space, i.e.,
\begin{equation}
\hat Q\hat U^{-1}\hat H\hat U\hat P=0.  
\end{equation}
Now the effective two-body interaction of Hermitian type can be determined in terms 
of unitary transformation $\hat U$ as
\begin{equation}
V_{\text{eff}}=V_{\text{low-k}}=\hat U^{-1}(\hat h_0-\hat v_{12})\hat U-\hat h_0,
\label{eq:Veff}
\end{equation}
where $\hat v_{12}$ stands for the bare NN interaction. The operator $\hat h_0$ denotes the one-body part 
of the two-body system and 
contains the kinetic energy of the interacting particles. It is important to notice
that in any case $\hat h_0$ commutes with 
the projection operators $\hat P$ and $\hat Q$.   
As it was shown by Suzuki \cite{Suzuki82}
 {the operator $\hat{U}$ is expressed as}
\begin{equation}
\hat U=(1+\hat\omega-\hat\omega^\dag)(1+\hat\omega\hat\omega^\dag+\hat\omega^\dag\hat\omega)^{-1/2},
\end{equation}
where an operator $\hat\omega$ fulfills relations $\hat\omega=\hat Q\hat\omega \hat P$ and 
$\hat\omega^2=\hat\omega^{\dag2}=0$. 
To evaluate the matrix elements of this operator $\hat\omega$ one should first solve 
the two-body eigenvalue equation
\begin{equation}
(\hat h_0+\hat v_{12})|\Phi_k\rangle=E_k|\Phi_k\rangle.  
\end{equation}
From eigenstates $|\Phi_k\rangle$ we determine those eigenstates $|\Phi_p\rangle$, which have 
the largest overlap 
with the $\hat P$ space. After the respective matrix elements of $\hat\omega$ and later $\hat U$ may be defined 
in terms of $\hat P(\hat Q)$ eigenstates.
This matrix element of $\hat U$ can then be used to determine the matrix elements of the effective 
interaction $V_{\text{eff}}$ in $\hat P$ space (for the details 
see \cite{Bozek06, Fujii04}). In this way, one obtains the effective Hamiltonian 
$\hat H_{\text{eff}}=\hat h_0+\hat V_{\text{eff}}$.
Diagonalising it in the low-momentum model space ($\hat P$ space), 
one obtains eigenvalues which are identical
to the diagonalization of the original Hamiltonian $\hat h_0+\hat V$ in the complete space.
Moreover, the solution of 
 {the}
Lippmann-Schwinger equation for NN scattering phase shifts using $V_{\text{low-k}}$
with a cutoff $\Lambda$ yields the same phase shifts as obtained from original 
interaction $\hat v_{12}$ without a cutoff. If the underlying interaction is a realistic interaction,
fitted to reproduce the experimental phase shifts below $\Lambda$,
these phase shifts will be also reproduced by $V_{\text{low-k}}$.

If the cutoff $\Lambda$ is chosen around $\Lambda$ = 2 fm$^{-1}$ the resulting $V_{\text{low-k}}$
is found to be essentially model independent, i.e., is independent on the underlying 
realistic interaction $\hat v_{12}$. 
In this sense $V_{\text{low-k}}$ is unique and, as it reproduces
the NN scattering phase shifts it can also be regarded as a realistic interaction as, e.g., 
the CD-Bonn \cite{Machleidt96} or Argonne V18 \cite{Wiringa95} potentials.
 
Originally $V_{\text{low-k}}$ is nonlocal and defined in terms of
matrix elements in a basis of NN states labeled by relative momentum for pairs of nucleons.  
Thus for each partial wave channel there exists a matrix,
which represents $V_{\text{low-k}}(k,k')$ on a mesh of $N$ discretized relative
momenta $k$ and $k'$ 
in the range $0\leq k,k'\leq\Lambda$. 
Since this matrix is real and symmetric with respect to $k,k'$ one can diagonalize it, 
so that, it can be written as
a sum of $N$ real eigenvalues multiplied with the respective eigenvectors
\begin{equation}
V_{\text{low-k}}(k,k')=\sum\limits^N_{i=1} a_i f^\ast_i(k)f_i(k'),
\label{eq:sep-form}
\end{equation}
where $N$ is the number of mesh-points and the dimension of $V_{\text{low-k}}$ matrix. The eigenvectors $f_i(k)$ satisfy 
the orthogonality relation
\begin{equation}
\frac{2}{\pi}\int^{\Lambda}_0 dk k^2 f_i(k) f_j(k) = \delta_{ij}.
\end{equation}
In the following we will omit the symbol of complex conjugation
because all eigenvectors are real. 
The last equality (\ref{eq:sep-form}) is nothing else but the general definition of a separable potential of the 
rank $N$. If the rank of the separable potential equals 
the dimension of the matrix $V_{\text{low-k}}(k,k')$
the whole information is exactly restored from the eigenvalues $a_i$ and eigenvectors $f_i$.
As we will see later, some of eigenvalues $a_i$ can be zero or negligibly small
so that one can reduce the rank of separable interaction taking into account 
only the $n$ eigenvalues with largest absolute values.
It leads to a new approximated separable interaction $V^{[n]}_{\text{low-k}}(k,k')$
\begin{equation}
V_{\text{low-k}}(k,k')\simeq V^{[n]}_{\text{low-k}}(k,k')=\sum\limits^n_{i=1} a_i f_i(k)f_i(k'), \hspace{1cm} (n\le N).
\label{eq:sep-fit-form}
\end{equation} 
The low-rank separable representation of NN interaction leads to significant simplifications in 
many-body calculations.

\begin{figure}

\begin{center}

\mbox{ \includegraphics[width = 8.5cm]{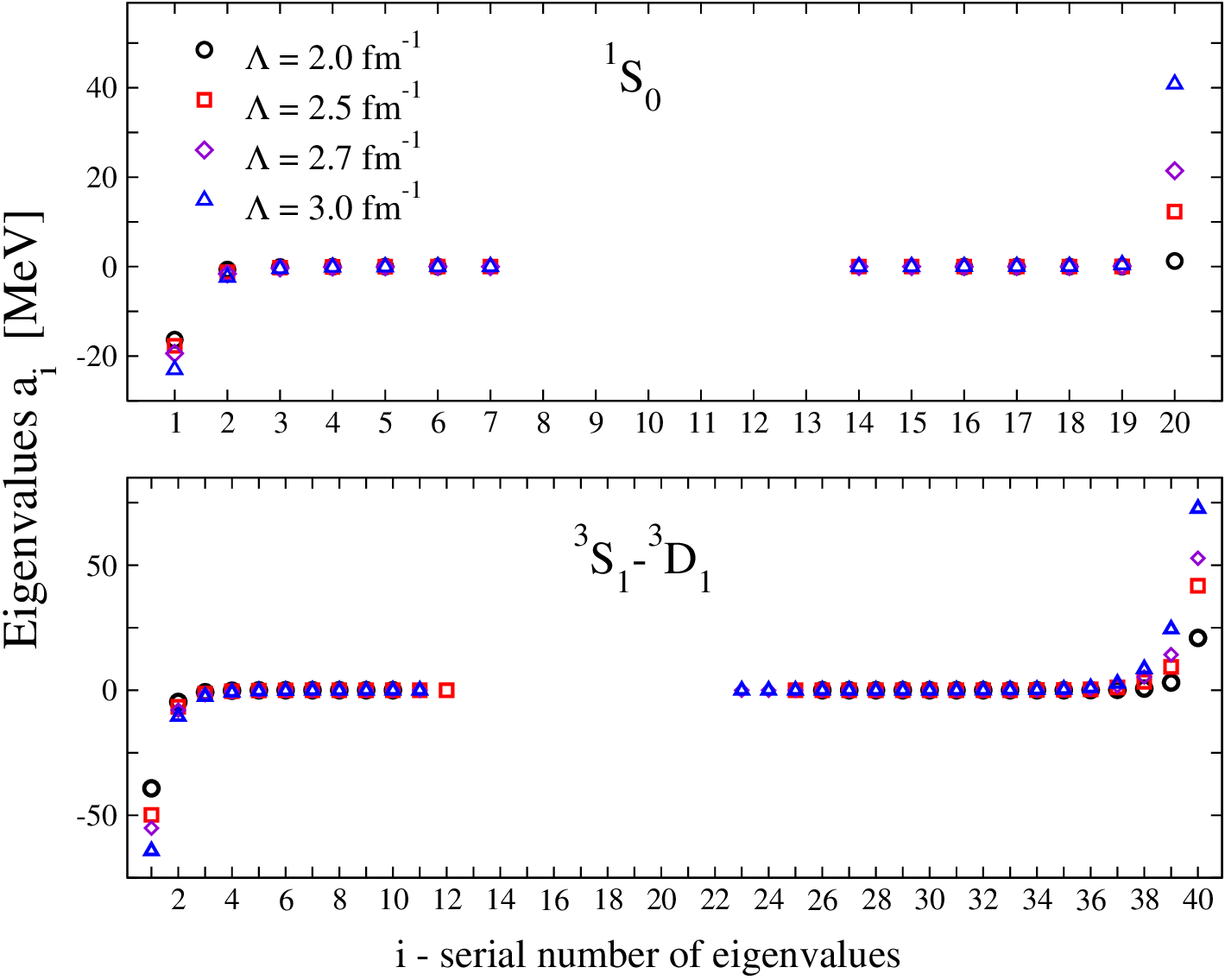} }

\end{center}

\caption{(Color online) 
Top: Nonzero eigenvalues $a_i$ of $^1S_0$ channel.
Bottom: Nonzero eigenvalues $a_i$ of $^3S_1$-$^3D_1$ channel.}
\label{fig:1s0_eigenvalues}

\end{figure}


The effective interaction $V_{\text{low-k}}$ as well as its separable form is nonlocal 
and defined in terms
of matrix elements in momentum space.
It implies that the HF calculations has to be performed in a Hilbert space
using an appropriate basis $\mid \alpha \rangle$, $\mid \beta \rangle$, $\dots$
The HF Hamiltonian is then expressed in terms of the matrix elements between these
basis states $\langle \alpha \mid H_{\text{HF}}\mid \beta\rangle$ and the HF 
single-particle (s.-p.)
states $\mid \Psi_n \rangle$ are expressed through the expansion coefficients in the basis
\begin{equation} 
|\Psi_n\rangle=\sum\limits_\alpha|\alpha\rangle\langle\alpha|\Psi_n\rangle
=\sum\limits_\alpha c_{n\alpha}|\alpha\rangle.
\end{equation}
The part of the HF Hamiltonian originating from $V_{\text{low-k}}$ 
can be expressed in terms of two-body matrix elements by
\begin{equation}
\langle \alpha \mid H_{\text{low-k}}|\beta\rangle
=\sum\limits_{\gamma,\delta}\langle\alpha\gamma|V_{\text{low-k}}|\beta\delta\rangle
\rho_{\gamma\delta},
\end{equation}
where $\rho_{\gamma\delta}$ is the single-particle density matrix.
In order to investigate the bulk properties of finite nuclei we perform HF calculations
within the spherical Wigner-Seitz cell assuming a plane wave single-particle
basis \cite{Montani04,vanDalen09}.

\section{Results and Discussion}
\label{sec:Results}
In the following we discuss results for symmetric nuclear matter as well as finite nuclei
obtained from HF calculations. These calculations are performed in the model space,
which is defined by a cutoff parameter $\Lambda$ in the two-body scattering equation,
employing the corresponding low-momentum interaction $V_{\text{low-k}}$, which is derived
from the CD-Bonn \cite{Machleidt96} interaction using the technique described in Sec.\ref{sec:NN-interaction}.
The NN interaction has been restricted to partial waves with total angular momentum $J$ 
less equal 6. 

\begin{figure}

\begin{center}

\mbox{ \includegraphics[width = 8.5cm]{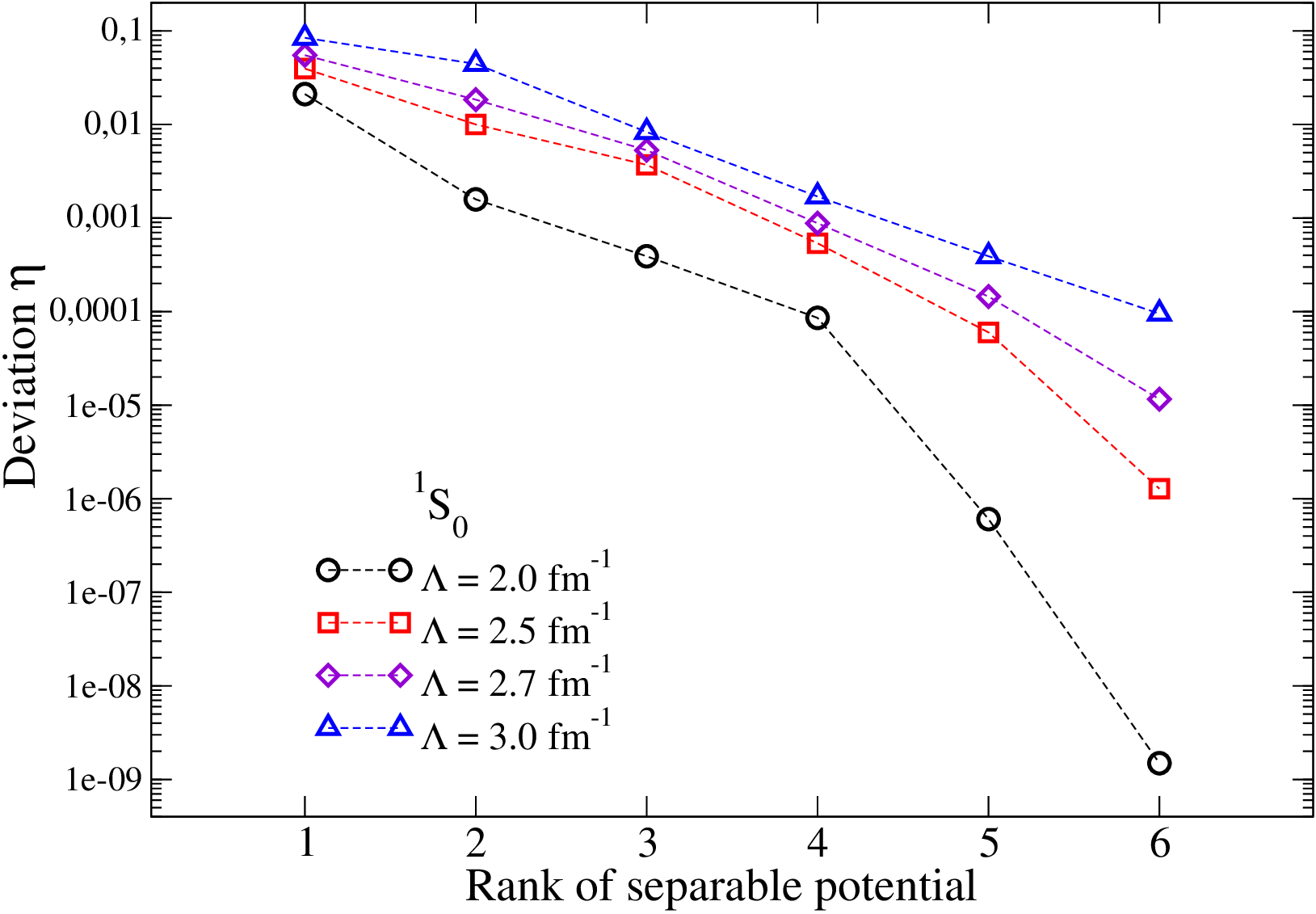} }

\end{center}

\caption{(Color online) Squared deviation of the separable $V^{[n]}_{\text{low-k}}(k,k')$ 
from the original $V_{\text{low-k}}(k,k')$
 in $^1S_0$ channel for different values of the cutoff parameter $\Lambda$.}
\label{fig:deviation1s0}

\end{figure}

We start our discussion with the comparison of the eigenvalues $a_i$ 
obtained from diagonalization of 
$20\times20$ matrix of $V_{\text{low-k}}(k,k')$ in $^1S_0$ channel.
The resulted nonzero eigenvalues are shown on the top of Fig.\ref{fig:1s0_eigenvalues}
for different values of $\Lambda$. As it was discussed above, $V_{\text{low-k}}$ interaction becomes
model independent at $\Lambda$ = 2 fm$^{-1}$. At this value of the cutoff parameter $\Lambda$
the diagonalization procedure yields only 11 nonzero eigenvalues, other words, 
$V_{\text{low-k}}$ interaction in $^1S_0$ channel is a separable interaction
of the 11$th$ rank or, 
following (\ref{eq:sep-form}), one can write
\begin{equation}
V^{[11]}_{\text{low-k}}(k,k')=V_{\text{low-k}}(k,k').
\end{equation}
The nonzero eigenvalues are essentially independent on $N$, the dimension of the
matrix representing $V_{\text{low-k}}$. 
Going further one can notice, that many of the nonzero eigenvalues are nevertheless very small,
and only some of them, e.g., at $i=1,2,20$ carry the main part of the information about
the interaction model. This gives rise to a substantial lowering of the rank of separable potential, 
as it was shown in Eq.(\ref{eq:sep-fit-form}).
With the increase of the cutoff $\Lambda$ the absolute values of the eigenvalues 
increase as well
and as a consequence the rank $n$ of the separable form $V^{[n]}_{\text{low-k}}$ 
defined in (\ref{eq:sep-fit-form}) has to be increased to achieve a reasonable
accuracy. Increasing $\Lambda$ more information about the short-range components
of the underlying bare interaction is included, which requires a larger rank in
the separable representation of the interaction.

In case of the coupled channels, like $^3S_1$-$^3D_1$ channel, the dimension $N$
of the 
$V_{\text{low-k}}$ matrix is twice as large if one keeps the number of
mesh-points in each channel the same as for the uncoupled partial waves. It turns
out that also the number of 
nonzero eigenvalues increases as shown in the lower panel of Fig.\ref{fig:1s0_eigenvalues}.
It is obvious that the rank of the separable potential should be higher than for $^1S_0$ channel.
It is a general feature that coupled channels require higher rank separable interaction than 
the uncoupled ones \cite{Haidenbauer84}.
Also one observes pairs of positive and negative eigenvalues which have about
the same absolute value.
This picture remains  for higher values of $\Lambda$. As we will see later, this symmetry
between positive and negative eigenvalues will play a crucial role in the
convergence of the separable form $V^{[n]}_{\text{low-k}}$ to the initial $V_{\text{low-k}}$  
with increase of rank.

In order to determine a minimal rank for a reliable separable approximation in 
each channel we calculate the square deviation $\eta$ of the separable form $V^{[n]}_{\text{low-k}}$ from the original 
potential $V_{\text{low-k}}$ for each rank $n$
\begin{equation}
\eta=\sum_{k,k'}\left|V_{\text{low-k}}(k,k')-V^{[n]}_{\text{low-k}}(k,k')\right|^2
/\sum_{k,k'}\left| V_{\text{low-k}}(k,k')\right|^2.
\end{equation}
 {In Fig.\ref{fig:deviation1s0} the deviation for $^1S_0$ channel 
at different values of the cutoff $\Lambda$ is shown.  
At $\Lambda$ = 2 fm$^{-1}$ one observes a fast convergence to zero deviation already at the rank $n=2$.
The growth of the cutoff monotonically increases the rank of the separable potential.
At $\Lambda$ = 3 fm$^{-1}$ one may expect a good accuracy starting from $n=5$.}


\begin{figure}
\begin{center}
\mbox{ \includegraphics[width = 8.5cm]{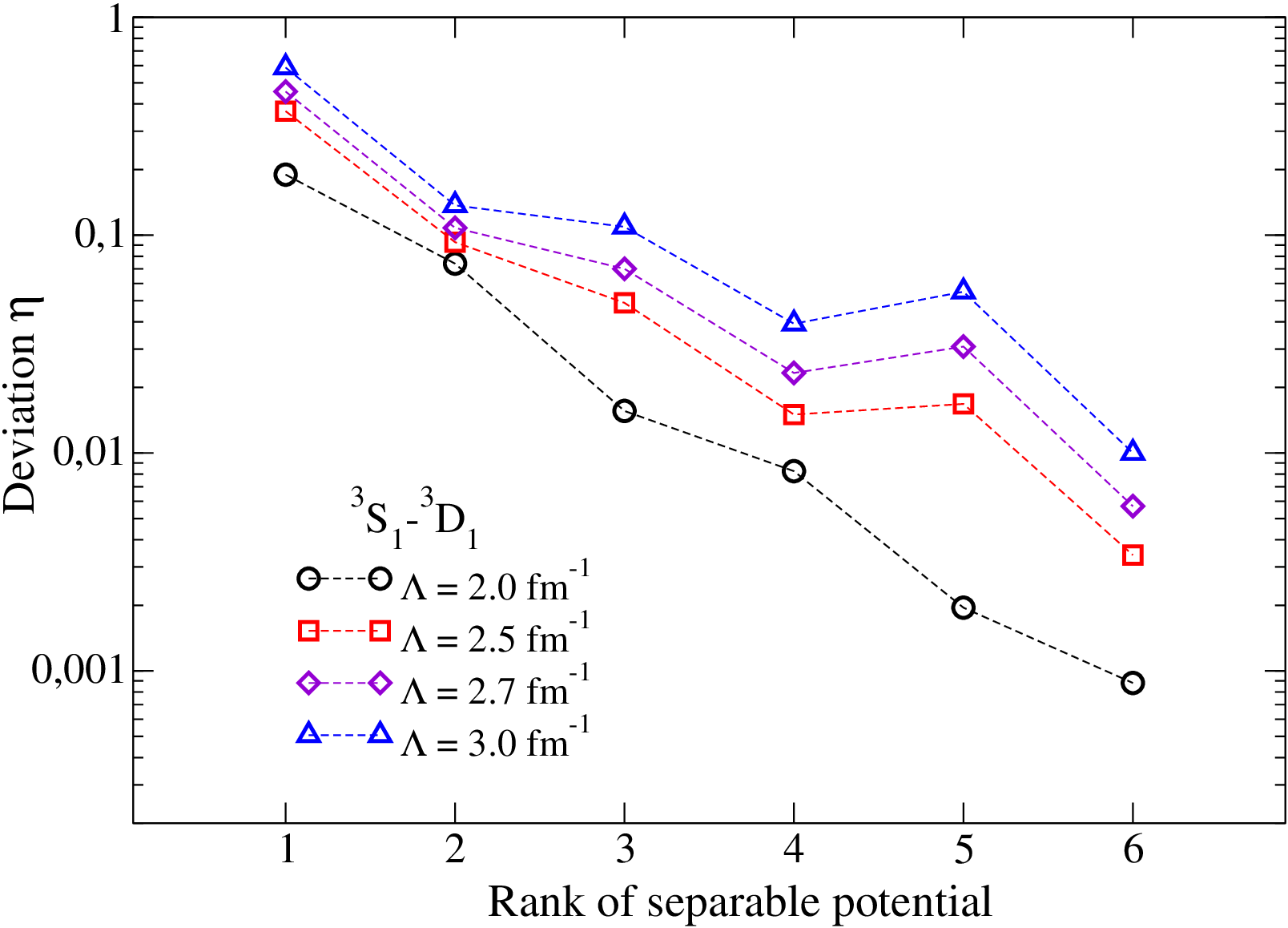} }
\end{center}
\caption{(Color online) Squared deviation of the separable $V^{[n]}_{\text{low-k}}(k,k')$ 
from the original $V_{\text{low-k}}(k,k')$
in $^3S_1$-$^{3}D_1$ channel for different values of the cutoff parameter $\Lambda$.}
\label{fig:deviation3s1}
\end{figure}

The deviation $\eta$ for $^3S_1$-$^3D_1$ channel is displayed in Fig.\ref{fig:deviation3s1}. 
First, at low $n$ 
the absolute value of the deviation is one order of magnitude higher than for uncoupled $^1S_0$
channel. Increasing the rank one observes a 
non-monotonic, oscillating decrease of $\eta$, specially for high $\Lambda$.
As we have seen, the diagonalization of the channel $^3S_1$-$^3D_1$ yields
both positive 
and negative eigenvalues, which are symmetrically distributed over $i$.
So that they form "pairs" with very similar absolute values.
Assuming the odd rank we take into account either
uncompensated positive or negative eigenvalue. 
This eigenvalue will be compensated in the next (even) rank, and the
accuracy will be significantly improved.

\begin{figure}
\begin{center}
\mbox{ \includegraphics[width = 8.5cm]{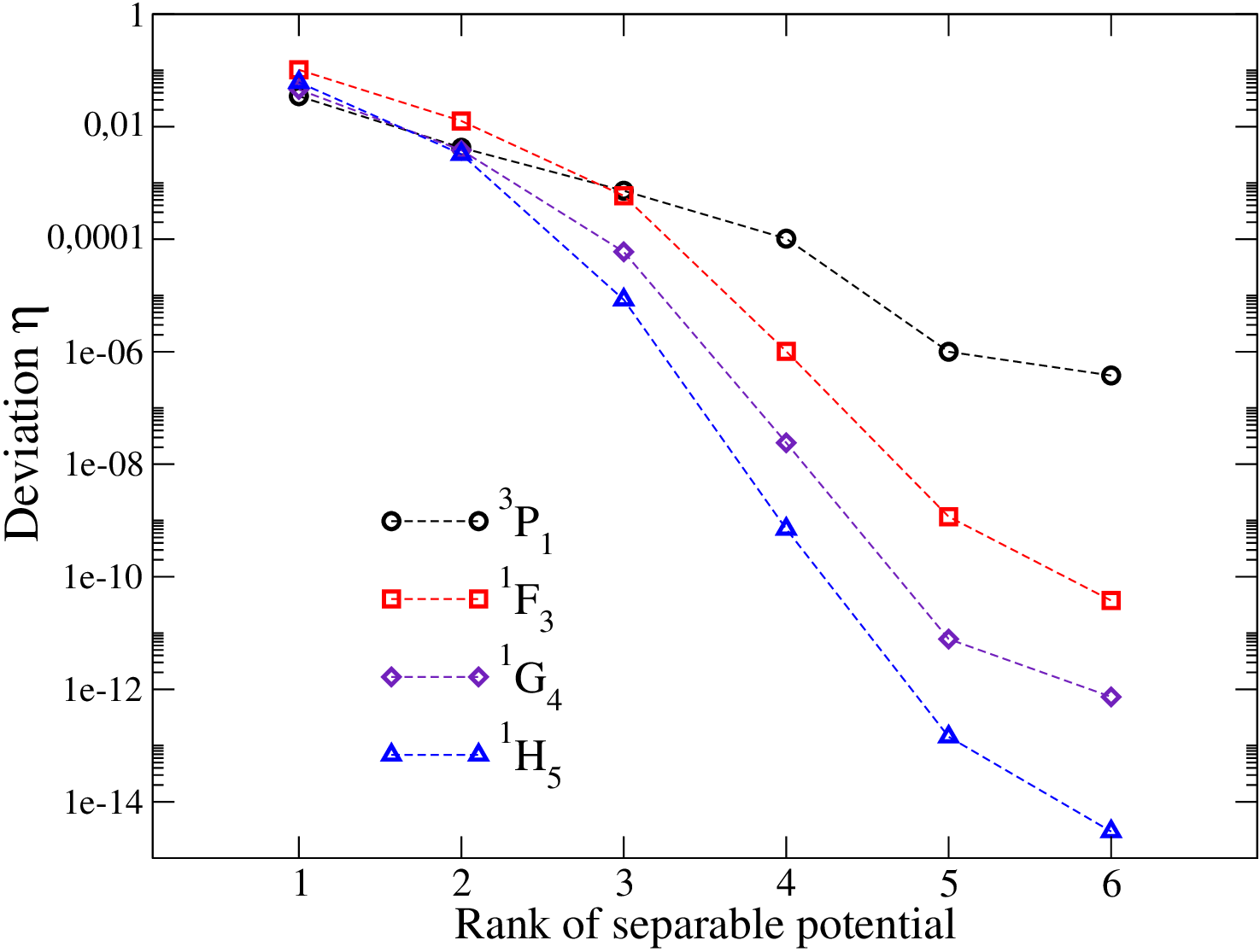} }
\end{center}
\caption{\label{fig:deviation_uncoupl} (Color online) Squared deviation of the separable $V^{[n]}_{\text{low-k}}(k,k')$ 
from the original $V_{\text{low-k}}(k,k')$
for various uncoupled channels at $\Lambda$ = 2 fm$^{-1}$.}
\end{figure}

\begin{figure}
\begin{center}
\mbox{ \includegraphics[width = 8.5cm]{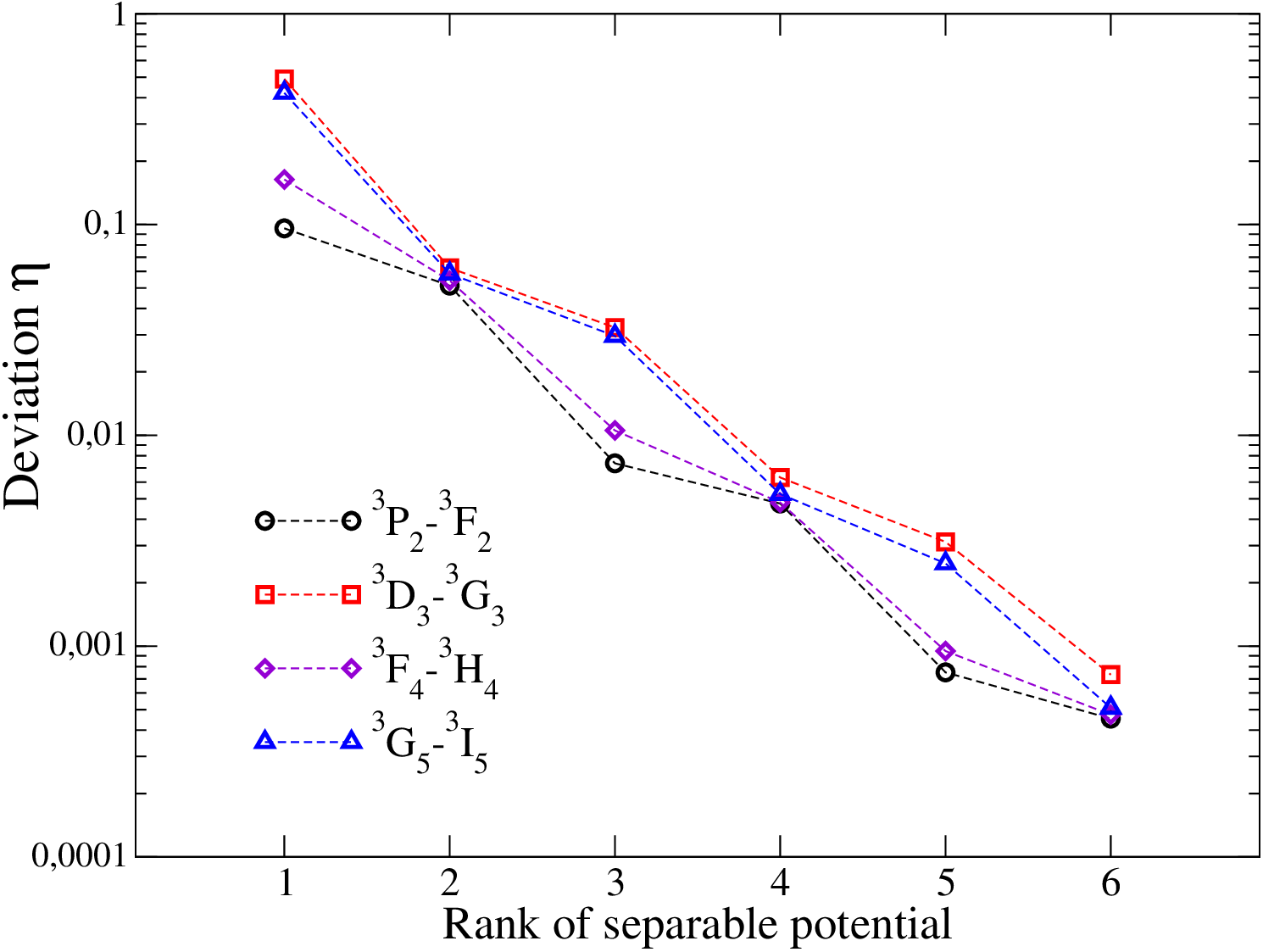} }
\end{center}
\caption{\label{fig:deviation_coupl} (Color online) Squared deviation of the separable $V^{[n]}_{\text{low-k}}(k,k')$ 
from the original $V_{\text{low-k}}(k,k')$
for various coupled channels at $\Lambda$ = 2 fm$^{-1}$.}
\end{figure}

 {The deviation $\eta$ for various other channels at $\Lambda$ = 2 fm$^{-1}$ is shown in Figs.\ref{fig:deviation_uncoupl},~\ref{fig:deviation_coupl}.
In the following we choose the second rank approximation for the uncoupled channels ($n=2$) and 
the third rank for the coupled one ($n=3$). Below, the respective separable version of $V_{\text{low-k}}$ 
will be referred to as $V^{[2,3]}_{\text{low-k}}$.}
 
\begin{figure}
\begin{center}
\mbox{ \includegraphics[width = 8.5cm]{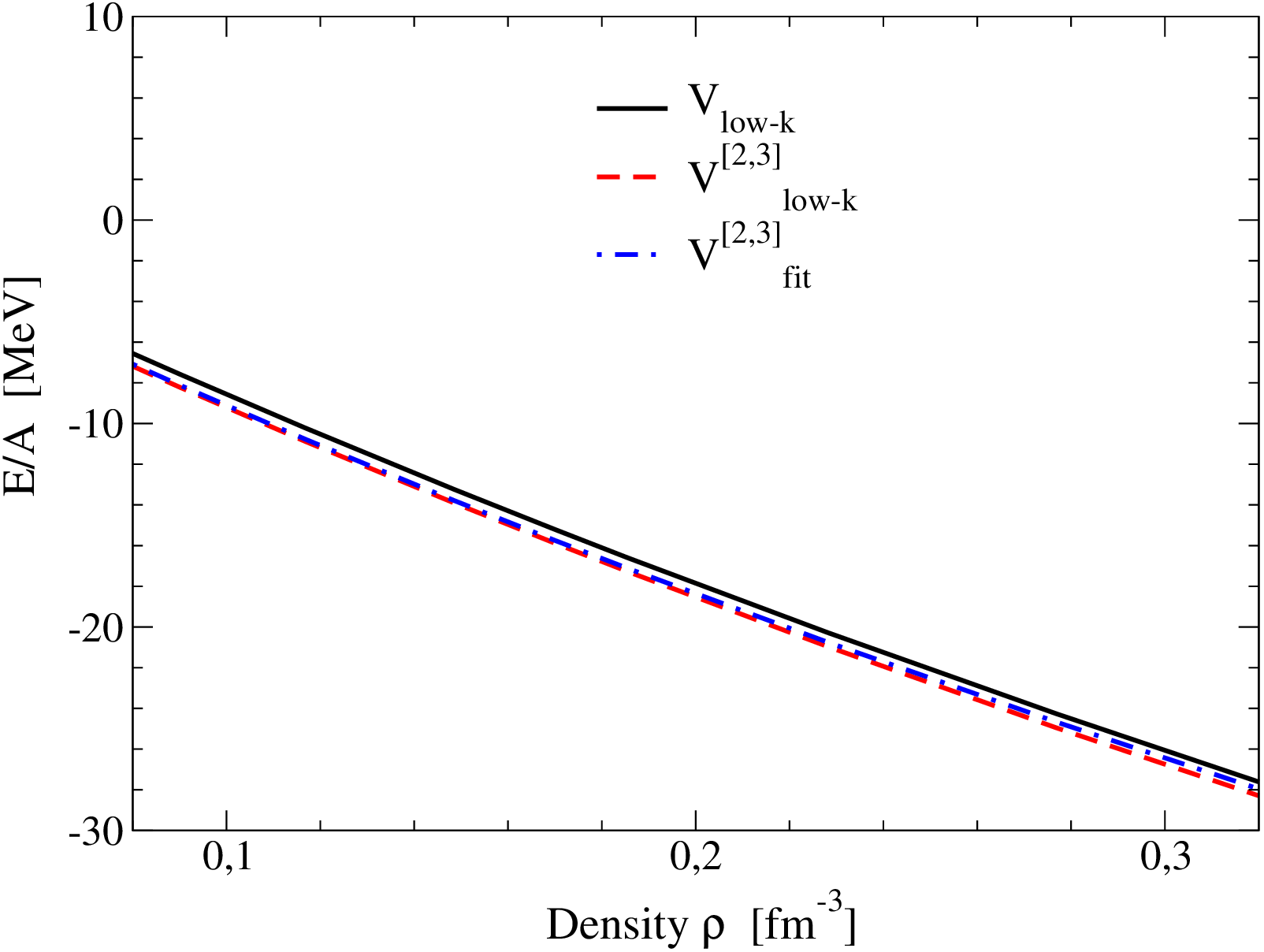} }
\end{center}
\caption{ \label{fig:SM} (Color online) Energy per nucleon of symmetric nuclear matter as a function of 
the density. Results of $V_{\text{low-k}}$
interaction (solid line) compared with the separable form $V^{[2,3]}_{\text{low-k}}$ 
(dashed line) and the respective fitted form $V^{[2,3]}_{\text{fit}}$ (dashed-dotted line).}
\end{figure}

Now let us turn to the binding energy of symmetric nuclear matter, which is displayed in 
Fig. \ref{fig:SM}. The HF calculations using $V^{[2,3]}_{\text{low-k}}$ (dashed
line) yields essentially the same result as the one employing the original
$V_{\text{low-k}}$ interaction (solid curve). The deviation does not exceed $1\%$ at the saturation density $\rho_0$ and $1.7\%$
at the density $2\rho_0$. We also compared the binding energy of pure neutron matter for both potentials
and found that the discrepancy is less than $1\%$ for the same range of densities.

However, neither of the 2 calculations yields a
saturation point, i.e. a minimum in the energy versus density plot, as it has 
been observed before \cite{Kuckei03,Bogner05}.
This absence of the saturation is one of the main problems in calculations of nuclear matter
employing $V_{\text{low-k}}$. It cannot be cured by the inclusion of correlations beyond
the HF approximation, e.g., by means of the BHF approximation \cite{Bozek06}. 
Recent relativistic calculations by van Dalen and M\"uther demonstrate 
that saturation can be achieved within the $V_{\text{low-k}}$ approach
by inclusion of relativistic effects in dressing the Dirac spinors which are
used to evaluate the underlying realistic interaction \cite{vDM09}.

All the results obtained so far indicate
that $V^{[2,3]}_{\text{low-k}}$ is an accurate low-rank separable
representation of $V_{\text{low-k}}$ interaction.
However, in order to make it accessible to other users, it should be parameterized
in a simple form.
Here we suggest the fitting function for all 
$f_i(k)$ in all channels
\begin{equation}
f_i(k)=\alpha_i+(\beta_i \exp{(\gamma_i k^{\delta_i})}+\mu_i)\sin(k\sigma_i  + \lambda_i),
\label{eq:fit}
\end{equation}
which contains 7 parameters for each partial wave channel and each $f_i(k)$.
In the Table \ref{tab:uncoupled}, we summarized all parameters of the separable fitted form 
for uncoupled channels, while all parameters for the coupled channels
are shown in the Table \ref{tab:coupled}. 
By using the values from both tables one can reproduce the fitted
version of $V^{[2,3]}_{\text{low-k}}$ for a given partial wave channel.
In the following we will identify the respective separable fitted potential
as $V^{[2,3]}_{\text{fit}}$.

In order to check the accuracy of our fit we perform HF calculations
of nuclear matter 
employing $V^{[2,3]}_{\text{fit}}$. 
The respective binding energy
as a function of the density of symmetric nuclear matter are displayed 
on Fig.\ref{fig:SM}
by a dashed-dotted line. One observes that at up to saturation density
$\rho_0\simeq0.16$ fm$^{-3}$ the fitted potential $V^{[2,3]}_{\text{fit}}$
reproduces the results of $V^{[2,3]}_{\text{low-k}}$ (red dashed), while
at higher densities it becomes slightly less bound and lies 
closer to the original $V_{\text{low-k}}$ (solid).
Thus the deviation of the fitted separable potential $V^{[2,3]}_{\text{fit}}$
from $V_{\text{low-k}}$ does not exceed $1\%$ of binding energy.
Not going into details we mention that the deviation rises mainly
from $^3S_1$-$^3D_1$ and $^3P_2$-$^3F_2$ coupled channels. 

\begin{table}
\begin{center}
\begin{tabular}{l c c c}
\hline
\hline
\ Interaction \ & \ $t_0$ [MeV fm$^3$] \ & \ $t_3$ [MeV fm$^{3+3\alpha}$] \ & \ \ $x_3$ \ \ \\
\hline
CT& -584.1 & 8330.7 & -0.5 \\
CT1& -548.0 & 7890.13 & -0.5 \\
CT2& -565.467 & 8180.0 & -0.5 \\
\hline
\hline
\end{tabular}
\end{center}
\caption{Parameters of the contact interaction
defined 
in Eq.(\ref{eq:contact-term}). The set CT was produced for $V_{\text{low-k}}$ \cite{vanDalen09},
while CT1 and CT2 supply $V^{[2,3]}_{\text{fit}}$.  }
\label{table:ct}
\end{table}


As we have already seen from Fig.\ref{fig:SM} $V_{\text{low-k}}$ interaction 
as well as its separable form $V^{[2,3]}_{\text{fit}}$
does not describe the empirical saturation point. 
In order to achieve the saturation in nuclear matter one has to add
three-body interaction terms or a density-dependent two-nucleon interaction.
Therefore we support the low-momentum interaction by a simple contact
interaction, which have been chosen following the notation of
the Skyrme interaction~\cite{Skyrme59,Vautherin1972}
\begin{equation}  
\Delta\mathcal{\nu}=\Delta\mathcal{\nu}_0+\Delta\mathcal{\nu}_3,
\label{eq:contact-term}
\end{equation}
with
\begin{equation}  
\Delta\mathcal{\nu}_0=\frac{1}{4}t_0\left[(2+x_0)\rho^2-(2x_0+1)(\rho_n^2+\rho_p^2)\right]
\end{equation}
and
\begin{equation}  
\Delta\mathcal{\nu}_3=\frac{1}{24}t_3\rho^\alpha\left[(2+x_3)\rho^2-(2x_3+1)(\rho_n^2+\rho_p^2)\right],
\end{equation}
where $\rho_p$ and $\rho_n$ are the local densities of nucleons 
while the total matter density
is denoted as $\rho=\rho_p+\rho_n$. The values of $\alpha$ and $x_0$ were 
fixed at $\alpha=0.5$, $x_0=0.0$,
while $t_0$, $t_3$, $x_3$ were fitted in such a way that HF calculations 
using $V_{\text{low-k}}$ or $V^{[2,3]}_{\text{fit}}$
plus the contact term (\ref{eq:contact-term}) reproduces both the empirical 
saturation point of the symmetric nuclear
matter and the symmetry energy at saturation density.
Following \cite{vanDalen09} the contact interaction 
produced for $V_{\text{low-k}}$ will be labeled by CT, and the respective
interaction model $V_{\text{low-k}}$+CT. 
For the fitted potential $V^{[2,3]}_{\text{fit}}$
we suggest two possible parameterizations: CT1 and CT2.
Their parameters and properties of nuclear matter
are shown in Tables \ref{table:ct} and \ref{table:ct_bulk}, respectively.

\begin{figure}
\begin{center}
\mbox{ \includegraphics[width = 8.5cm]{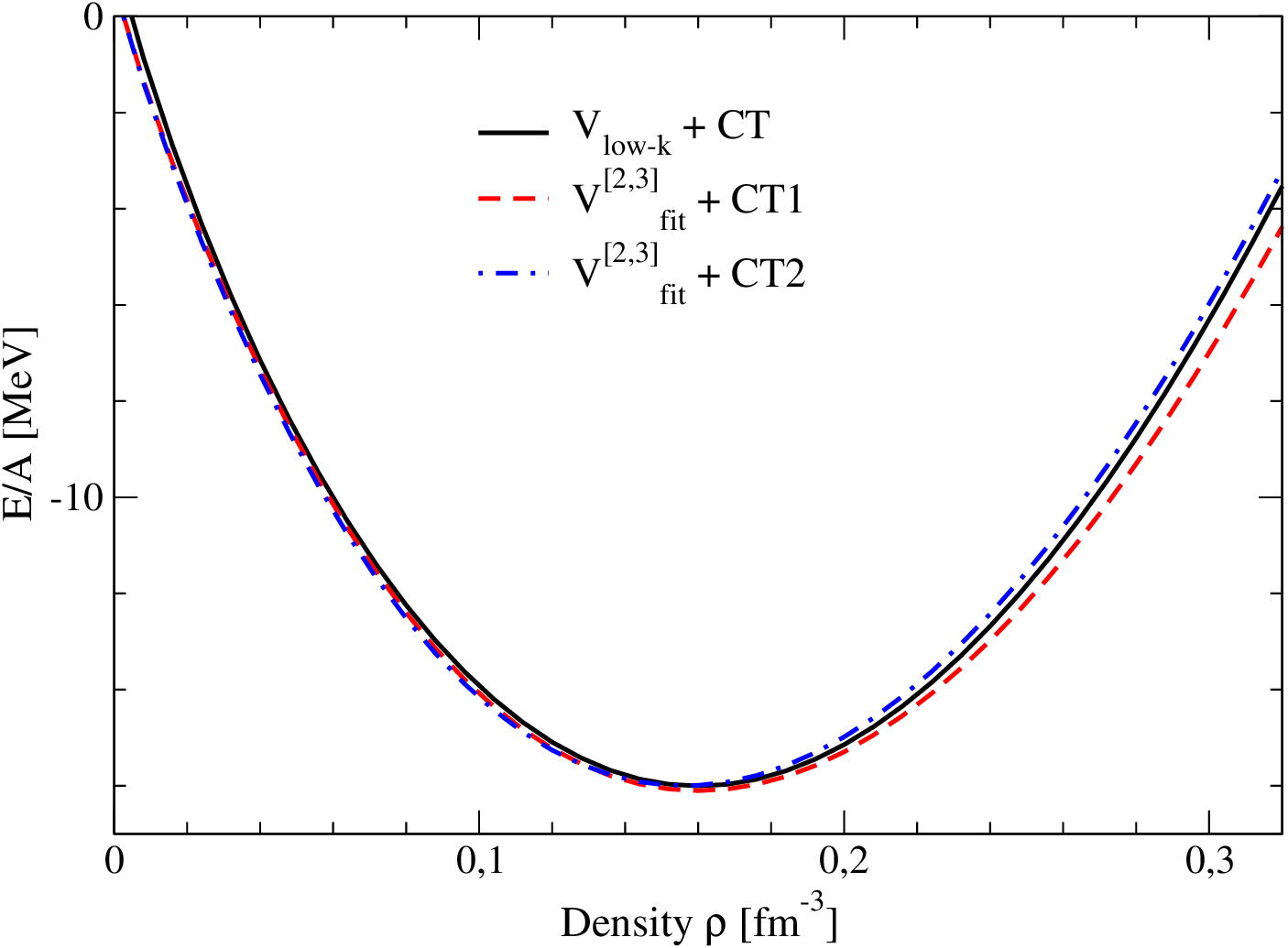} }
\end{center}
\caption{\label{fig:SM_ct}(Color online) Energy per nucleon of symmetric nuclear matter 
as a function of the density. 
Results of $V_{\text{low-k}}$ + CT
interaction (solid line) compared with the fitted separable form 
$V^{[2,3]}_{\text{fit}}$ + CT1 (dashed line) and 
$V^{[2,3]}_{\text{fit}}$ + CT2 (dashed-dotted line).}
\end{figure}

The interaction
$V^{[2,3]}_{\text{fit}}$ + CT1 
gives the binding energy per nucleon of symmetric nuclear matter 
E/A=-16.1 MeV at the density $\rho_0=0.16$ fm$^{-3}$. 
The HF calculations of nuclear matter (see Fig.\ref{fig:SM_ct})
for $V^{[2,3]}_{\text{fit}}$ + CT1 
give results (dashed line) very similar to 
the non-separable initial interaction 
$V_{\text{low-k}}$ + CT (solid line).
However, in calculation of finite nuclei we observe a deviation of about 0.12MeV 
in the binding energy of light nuclei, like $^{16}$O 
(see Table \ref{table:nuclei}).
The picture can be improved if we
assume, that the saturation density is not defined exactly and allow for
a small deviation. Along this line the second parameterization CT2
was produced. The interaction $V^{[2,3]}_{\text{fit}}$ + CT2
gives E/A=-16.0 MeV at the density $\rho_0=0.156$ fm$^{-3}$. 
This corresponds to a small shift of the saturation point
with respect to the initial $V_{\text{low-k}}$ interaction
(see Fig.\ref{fig:SM_ct}).
It allows us to improve the accuracy in 
the binding energies of finite nuclei:
one can notice that the contact interaction 
CT2 leads to a better description than CT1.
However, comparing the rms charge radii of nuclei in 
Table \ref{table:nuclei}, we see
that due to the shift in saturation density the interaction 
$V^{[2,3]}_{\text{fit}}$ + CT2 yields larger radii than the interaction 
$V^{[2,3]}_{\text{fit}}$ + CT1.

For all models considered here the compressibility modulus
at saturation density is in the range $240.5\leq K \leq 258$ MeV.
This means that the respective equations of state displayed in Fig.\ref{fig:SM_ct} are rather soft,
at least at densities up to about two times saturation density.
Such a prediction of a soft EoS is in agreement with 
data extracted from heavy ion reactions. For example, heavy 
ion data for transverse flow \cite{Stoicea04} 
or from kaon production \cite{Sturm01} support
the picture of a soft EoS in symmetric nuclear matter.
 {This value for the compressibility modulus is also
in agreement with that of the Skyrme interaction which reproduce
correctly the breathing mode in nuclei (giant isoscalar 
resonance)~\cite{Shlomo2006}.}

\begin{table}
\begin{center}
\begin{tabular}{l c c c}
\hline
\hline
\ Interaction \ & \ $\rho_0$ [fm$^{-3}$] \ & \ E/A$(\rho_0)$ [MeV] \ & \ K [MeV] \ \\
\hline
$V_{\text{low-k}}$+CT        & 0.16  & -16.0 & 258 \\
$V^{[2,3]}_{\text{fit}}$+CT1 & 0.16  & -16.1 & 241.9 \\
$V^{[2,3]}_{\text{fit}}$+CT2 & 0.156 & -16.0 & 240.5 \\
\hline
\hline
\end{tabular}
\end{center}
\caption{Bulk properties of symmetric nuclear matter derived from 
$V_{\text{low-k}}$ and its separable representation. 
They are supplemented by the respective contact interaction.}
\label{table:ct_bulk}
\end{table}



\begin{table}
\begin{center}
\begin{tabular}{ l c c c c c}
\hline
\hline
Interaction & \ \ $^{16}$O \ & \ $^{40}$Ca \ & \ $^{48}$Ca \ & \ $^{60}$Ca & \ $^{208}$Pb \ \\
\hline
 &\multicolumn{1}{c}{ E/A [MeV] }\\
\cline{2-6}
$V_{\text{low-k}}$+CT       & -7.91 & -8.57 & -8.42  & -7.75 & -7.76\\
$V^{[2,3]}_{\text{fit}}$+CT1& -7.79 & -8.56 & -8.35  & -8.78 & -7.76\\
$V^{[2,3]}_{\text{fit}}$+CT2& -7.84 & -8.58 & -8.37  & -8.79 & -7.76\\
Experiment                  & -7.98 & -8.55 & -8.67  &  --   & -7.87\\
 &\multicolumn{1}{c}{  $r_{ch}$ [fm] }\\
\cline{2-6}
$V_{\text{low-k}}$+CT       &  2.79 &  3.50 &  3.54  &  3.68 &  5.51\\
$V^{[2,3]}_{\text{fit}}$+CT1&  2.81 &  3.51 &  3.55  &  3.68 &  5.52\\
$V^{[2,3]}_{\text{fit}}$+CT2&  2.82 &  3.53 &  3.58  &  3.71 &  5.56\\
Experiment                  &  2.74 &  3.48 &  3.47  &  --   &  5.50\\
\hline
\hline
\end{tabular}
\end{center}
\caption{\label{table:nuclei}
The binding energy per nucleon and rms charge radii of finite nuclei.
Experimental data taken from Refs. \cite{Brown98}-\cite{Fricke95}.}
\end{table}

\section{Summary and Conclusion}

In the last decade it has become popular to perform nuclear structure 
calculations using the low-momentum NN interaction 
$V_{\text{low-k}}$ (see for instance the recent Ref.~\cite{Lesinski2009,Hebeler2009}).
This interaction is constructed from a realistic NN interaction
by introducing a cutoff $\Lambda$ in the relative momenta 
of the interacting nucleons. We used a model space technique
on the base of the unitary model operator approach to
separate the low-momentum and high-momentum parts of
the initial CD-Bonn interaction. 
The cutoff parameter $\Lambda$ was fixed at
$\Lambda$ = 2 fm$^{-1}$ so that a $V_{\text{low-k}}$ is obtained, which is
essentially independent on the underlying bare NN interaction. 

The resulted $V_{\text{low-k}}$ interaction
is nonlocal and defined
in terms of matrix elements in momentum space
for each partial wave channel.
This allows us to use a diagonalization method in order 
to express the matrix elements in a 
separable form. We investigate the separability
in different channels with increase of the cutoff $\Lambda$.
It was found that at $\Lambda$ = 2 fm$^{-1}$
the low-momentum interaction can accurately be
approximated by a low-rank separable interaction. This separable interaction is
parameterized to make it accessible for other nuclear structure calculations.

A density dependent contact interaction is added to reproduce the saturation
property of infinite nuclear matter. HF calculations using this interaction
model also reproduce the bulk properties of finite nuclei with good accuracy.
We demonstrate that this new separable representation of the $V_{\text{low-k}}$
interaction, reproduces the results derived from the original $V_{\text{low-k}}$
with a high accuracy. 

This work has been supported by the European
Graduate School ``Hadrons in Vacuum, in Nuclei and Stars'' (Basel, Graz,
T\"{u}bingen) and a grant Mu 705/5-2 of the Deutsche Forschungsgemeinschaft
(DFG)
 {and by CompStar, a Research Networking Programme of the European Science Foundation.}

\begin{table}[h]
\tiny
\begin{center}
\begin{tabular}{cccccccccc}
\hline 
\hline
Channel & $i$ & $a_i$ & $\alpha_i $ & $\beta_i$ & $\gamma_i$ & $\delta_i$ & $\mu_i$ & $\sigma_i$ & $\lambda_i$\\
\hline
$^1S_0$& $1$ &    -0.16344E+02  &   0.10772E-03  &  -0.43234E-03 &    0.17650E+00  &   0.92610E+00 &    0.10316E-02 &
     0.99335E+00  &   0.14191E+01 \\
& $2$ &    -0.66770E+00  &  -0.20749E-03 &   -0.18295E-01 &    0.66232E-01   &  0.86815E+00  &   0.20236E-01 &
     0.17925E+01  &   0.11085E+01 \\
 $^1P_1$ & $1$ &  0.14569E+02  &  -0.36052E-02  &   0.12380E-02  &   0.30093E+00  &   0.15720E+01 &    0.24098E-02 &
     0.46996E+00 &    0.14196E+01\\
& $2$ &      0.15105E+01  &   0.41103E-04  &  -0.30056E-02  &   0.46402E-01  &   0.12591E+01 &    0.36618E-02 &
     0.22676E+01 &   -0.66286E-02 \\

 $^3P_0$ & $1$ &     0.36518E+01  &   0.34915E-04  &  -0.39191E-03 &    0.74648E-01  &   0.35155E+01  &   0.35779E-03 &
     0.20292E+01  &   0.14199E+01 \\
& $2$ &   -0.36339E+01 &    0.12254E-03  &   0.20982E-03  &  -0.59888E+00  &   0.10769E+01  &  0.30373E-03 &
    -0.21878E+01  &   0.96652E+01 \\

 $^3P_1$ & $1$ &     0.15410E+02 &    -0.33598E-03  &   0.15182E-02 &   -0.12701E-02  &   0.54199E+01 &   -0.85450E-03 &
     0.59810E+00 &    0.52799E+00 \\
& $2$ &     0.11011E+01 &  -0.44421E-03  &   0.18896E-02  &  -0.63158E+00 &    0.24471E+01  &   0.94546E-04 &
     0.73498E+00  &   0.22558E+00 \\

$^1D_2$ & $1$ & -0.47228E+01  &   0.23948E-03  &   0.11514E-02  &   0.78532E-01  &   0.83080E+00  &  -0.14043E-02 &
     0.15502E+01  &   0.12639E+01 \\
& $2$ & -0.42720E+00  &   0.43276E-03 &   -0.13308E-02  &  -0.76108E+00  &   0.22797E+01  &   0.89215E-03 &
     0.12810E+01  &   0.17046E+01 \\
$^3D_2$ & $1$ & -0.14755E+02  &   0.23661E-03  &  -0.16214E-03  &  -0.20028E+00   &  0.47727E+01  &  -0.79636E-04 &
     0.12915E+01  &   0.17118E+01 \\
& $2$ & -0.20625E+01  &   0.46873E-03  &  -0.13299E-02  &  -0.70797E+00  &   0.22075E+01   &  0.85842E-03 &
     0.14395E+01  &   0.16497E+01 \\
$^1F_3$ & $1$ & 0.21276E+01   &  0.11682E-03  &  -0.24864E-03  &  -0.15328E+01  &  -0.76112E+00 &   -0.11725E-03 &
     0.20029E+01   &  0.14366E+01 \\
& $2$ & 0.45580E+00  &   0.45434E-03  &   0.34263E-02  &  -0.23846E+01  &  -0.12925E+01  &  -0.45576E-03 &
     0.18623E+01   &  0.14425E+01 \\

$^3F_3$ & $1$ & 0.11895E+01 &     0.12811E-03  &  -0.89537E-04  &  -0.63186E+00  &  -0.13835E+01 &   -0.12893E-03 &
     0.20960E+01  &   0.13909E+01 \\
& $2$ & 0.26740E+00  &   0.44076E-03  &   0.35397E-02  &  -0.24171E+01 &   -0.12481E+01 &   -0.44204E-03 &
     0.18298E+01  &   0.14500E+01 \\

$^1G_4$ & $1$ & -0.56713E+00  &   0.93511E-04  &  -0.32673E-03  &  -0.14289E+01 &   -0.10215E+01  &  -0.94613E-04 &
     0.19148E+01  &   0.13505E+01 \\
& $2$ & -0.09176E+00  &   0.92267E-02 &   -0.14963E-01  &  -0.55660E+01  &  -0.18735E+01  &  -0.92335E-02 &
    -0.37857E+00  &   0.16284E+01 \\
$^3G_4$ & $1$ & -0.30270E+01   &  0.10774E-03  &  -0.16121E-03  &  -0.77810E+00  &  -0.15648E+01  &  -0.10940E-03 &
     0.18745E+01  &   0.13214E+01 \\
& $2$ & -0.5061E+00  &   0.10191E-03  &  -0.62066E-03  &  -0.10817E+01  &  -0.10648E+01  &  -0.10392E-03 &
     0.31194E+01  &   0.13385E+01 \\
$^1H_5$ & $1$ & 0.06095E+01  &   0.14458E-06  &  -0.43467E-03  &  -0.30481E+00  &   0.34986E+01  &   0.43536E-03 &
     0.13486E+01  &  -0.29605E+00 \\
& $2$ & 0.01455E+01 &    0.99849E-04 &   -0.38155E-03  &  -0.64624E+00  &  -0.24138E+01  &  -0.10803E-03 &
     0.33677E+01  &   0.95975E+00 \\
$^3H_5$ & $1$ & 0.03736E+01   &  0.17206E-06  &  -0.45636E-03  &  -0.28889E+00  &   0.34083E+01  &   0.45708E-03 &
     0.13780E+01  &  -0.31990E+00 \\
& $2$ & 0.00727E+00  &   0.86739E-04  &  -0.36173E-03  &  -0.88687E+00  &   0.29817E+01  &   0.45115E-03 &
     0.32408E+01  &  -0.20198E+01 \\
$^1I_6$ & $1$ & -0.01407E+01   &  0.18615E-04  &  -0.27820E-04  &   0.15265E+01  &   0.97282E+00  &   0.15276E-04 &
     0.20513E+01  &   0.15767E+01 \\
& $2$ & -0.00281E+01  &   0.63551E+00  &  -0.26703E+00  &   0.18734E+00  &   0.19309E+01  &  -0.36758E+00 &
     0.39862E+00  &   0.15760E+01 \\
$^3I_6$ & $1$ & -0.08297E+01  & -0.18713E-04  &  -0.30695E+00  &   0.31341E-03  &   0.21961E+01  &   0.30694E+00 &
     0.19374E+01  &   0.16207E+01 \\
& $2$ & -0.0153E+01   &  0.94460E-04  &  -0.30640E+00  &   0.53129E-03  &   0.14854E+01  &   0.30630E+00 &
     0.33132E+01  &   0.79134E+00 \\
\hline
\hline
\end{tabular}
\end{center}
\caption{\label{tab:uncoupled} Parameters of $V^{[2,3]}_{\text{fit}}$ for uncoupled channels. See Eq.(\ref{eq:fit}).}
\end{table}

\begin{table}
\tiny
\begin{center}
\begin{tabular}{cccccccccc}
\hline 
\hline
Channel & $i$ & $a_i$ & $\alpha_i $ & $\beta_i$ & $\gamma_i$ & $\delta_i$ & $\mu_i$ & $\sigma_i$ & $\lambda_i$\\
\hline
$^3S_1$ & $1$ & -0.39195E+02  &   0.14710E-03  &  -0.27882E-02  &   0.10166E+00  &   0.98761E+00  &   0.33148E-02 &
     0.73493E+00  &   0.91775E+00 \\
& $2$ &  0.20913E+02  &  -0.97562E-04   &  0.33292E+00  &   0.50609E-03  &   0.87732E+00  &  -0.33339E+00 &
     0.14274E+01  &   0.12444E+01 \\
& $3$ & -0.46417E+01 &   -0.21867E-04  &   0.31258E+00  &   0.17566E-02  &   0.78784E+00  &  -0.31368E+00 &
     0.18855E+01   &  0.11761E+01  \\
$^3D_1$ & $1$ & -0.39195E+02  &   0.90122E-04  &   0.31369E-05  &   0.38917E-03  &   0.12444E+02  &  -0.93123E-04 &
     0.13384E+01  &   0.15561E+01 \\
& $2$ & 0.20913E+02  &   0.11207E-03  &  -0.13705E-04  &   0.48296E+00  &   0.24592E+01  &  -0.97861E-04 &
     0.21818E+01  &   0.15329E+01 \\
& $3$ & -0.46417E+01 &  -0.22645E-03 &   -0.41235E+00  &   0.40973E-03  &   0.18283E+01   &  0.41258E+00 &
     0.15668E+01  &   0.15335E+01 \\
$^3P_2$ & $1$ &     -0.11712E+02  &  -0.12771E-01  &   0.74967E-02  &   0.29157E-01  &   0.83736E+00  &   0.53084E-02 &
     0.11992E+00  &   0.14904E+01 \\
& $2$ &  -0.21898E+01  &  -0.92101E-04  &   0.28219E+00  &   0.19554E-02  &   0.11980E+01  &  -0.28296E+00 &
     0.12014E+01  &  -0.12709E+00 \\
& $3$ &  0.17808E+01  &   0.14009E-03  &   0.34622E+00  &   0.13719E-02  &   0.12162E+01  &  -0.34694E+00 &
     0.12159E+01  &   0.29393E+01 \\
$^3F_2$ & $1$ & -0.11712E+02  &  -0.37975E-05  &  -0.26753E-09  &   0.11862E+02  &   0.11520E+00  &  -0.69054E-05 &
     0.10568E+01  &  -0.53109E+00 \\
& $2$ &  -0.21898E+01  &   0.71589E-04  &  -0.98497E-14  &   0.22438E+02  &   0.54423E-01  &  -0.69308E-04 &
    -0.23032E+01  &  -0.46525E+01\\
& $3$ &  0.17808E+01  &   0.24733E-03  &  -0.69363E-04   &  0.12429E+01   &  0.45015E+00   &  0.40433E-03 &
     0.11919E+01  &   0.40019E+01\\
$^3D_3$ & $1$ &  -0.60466E+01  &   0.18332E-03  &   0.14266E+00  &   0.41443E-03  &   0.17445E+01  &  -0.14284E+00 &
     0.21656E+01  &   0.15038E+01\\
& $2$ &  0.54641E+01  &  -0.27373E-03  &  -0.45681E+00  &   0.67993E-03  &   0.90916E+00  &   0.45713E+00 &
     0.18906E+01  &   0.99510E+00 \\
& $3$ &  0.91410E+00  &  -0.11606E-03  &   0.18114E+00  &  -0.13213E+00  &   0.12510E-02  &  -0.15901E+00 &
     0.33728E+01  &  -0.79666E+00 \\
$^3G_3$ & $1$ & -0.60466E+01  &   0.41921E-04  &  -0.65759E-01  &   0.79378E-03  &   0.18207E+01  &   0.65717E-01 &
     0.14240E+01  &   0.16750E+01 \\
& $2$ & 0.54641E+01  &   0.15808E-03  &  -0.16419E-03  &  -0.10345E+00  &  -0.99477E+00  &   0.28959E-03 &
     0.19522E+01 &   -0.25907E+01 \\
& $3$ & 0.91410E+00  &   0.78827E-04  &  -0.73989E-01  &  -0.17382E-02   &  0.15332E+01  &   0.74065E-01 &
     0.30794E+01  &  -0.18745E+01 \\
$^3F_4$ & $1$ & -0.16095E+01  &   0.18033E-03  &   0.37879E+00  &  -0.39428E-03  &   0.37390E+00  &  -0.37852E+00 &
     0.21928E+01  &  -0.23390E+01\\
& $2$ &  0.56160E+00   &  0.24168E-03  &  -0.11138E+01  &  -0.80823E-04  &   0.27458E+01  &   0.11135E+01 &
     0.19919E+01  &   0.13089E+01 \\
& $3$ &  -0.30360E+00  &  -0.32058E-03  &  -0.11873E+01  &   0.28972E-03  &   0.16371E+01  &   0.11877E+01 &
     0.20173E+01  &   0.11166E+01 \\
$^3H_4$ & $1$ & -0.16095E+01  &   0.30449E-07   &  0.28581E-01  &  -0.69621E+01   & -0.54467E+00  &   0.15057E-05 &
     0.10041E+01  &   0.30546E+01 \\
& $2$ &  0.56160E+00   &  0.17531E-04   &  0.22351E+00   &  0.34114E-03  &   0.19278E+01 &   -0.22349E+00 &
     0.16346E+01  &  -0.12499E+01\\
& $3$ & -0.30360E+00  &  -0.14910E-02  &   0.20920E+00  &   0.26023E-02   &  0.24400E+01  &  -0.20770E+00 &
     0.59266E+00  &   0.16380E+01 \\
$^3G_5$ & $1$ &  0.16137E+01  &  -0.93523E-04  &  -0.48952E+00  &  -0.88510E-04   &  0.10928E+01   &  0.48961E+00 &
     0.25801E+01  &   0.11545E+01 \\
& $2$ & -0.12501E+01  &   0.13111E-03  &  -0.13521E+00  &  -0.10757E+00  &   0.10162E-02  &   0.12130E+00 &
     0.28269E+01  &   0.70560E+00 \\
& $3$ &  0.30880E+00  &  -0.31841E-03  &  -0.53598E+00  &   0.61243E-03   &  0.18397E+01 &    0.53633E+00 &
     0.23654E+01  &   0.99154E+00 \\
$^3I_5$ & $1$ & 0.16137E+01  &   0.14023E-04   &  0.36754E+00  &  -0.14039E-03  &   0.25880E+01  &  -0.36755E+00 &
     0.17978E+01  &   0.15991E+01 \\
& $2s$ & -0.12501E+01  &   0.19169E-04  &   0.29512E+00  &   0.15469E-03  &   0.26251E+01  &  -0.29510E+00 &
     0.17038E+01  &  -0.16316E+01 \\
& $3$ & 0.30880E+00   &  0.38821E-04   &  0.36156E+00  &   0.35017E-03  &   0.19665E+01  &  -0.36153E+00 &
     0.30449E+01  &  -0.21756E+01 \\
$^3H_6$ & $1$ & -0.30540E+00  &  -0.65438E-03  &   0.65143E+00  &   0.56031E-03   &  0.23634E+01  &  -0.65075E+00 &
     0.48747E+00   &  0.18318E+01 \\
& $2$ &  0.20960E+00   &  0.61333E-04   &  0.25966E+00  &  -0.32893E-03  &   0.13102E+01  &  -0.25972E+00 &
     0.25754E+01 &    0.11963E+01 \\
& $3$ &  -0.63700E-01  &   0.25601E-03  &   0.39808E+00  &   0.29243E-03  &   0.32313E+01  &  -0.39836E+00 &
     0.22583E+01  &   0.10325E+01 \\
$^3K_6$ & $1$ & -0.30540E+00  &  -0.10269E-04   &  0.23504E+00  &   0.14785E-03   &  0.31131E+01  &  -0.23503E+00 &
     0.18245E+01  &   0.14457E+01 \\
& $2$ & 0.20960E+00  &   0.12093E-04  &  -0.36799E+00  &   0.11473E-03  &   0.30036E+01  &   0.36798E+00 &
     0.18060E+01   &  0.14804E+01 \\
& $3$ & -0.63700E-01  &  -0.19663E-04   &  0.26322E+00   &  0.39336E-03  &   0.23907E+01  &  -0.26320E+00 &
     0.29488E+01 &    0.91309E+00 \\
\hline
\hline
\end{tabular}
\end{center}
\caption{Parameters of $V^{[2,3]}_{\text{fit}}$ for coupled channels. See Eq.(\ref{eq:fit}).}
\label{tab:coupled}
\end{table}










\end{document}